\documentclass[9pt]{article}
\usepackage{spconf,amsmath,graphicx}

\usepackage{cite}
\usepackage{graphicx}
\usepackage{amsfonts} 
\usepackage{amsmath}
\usepackage{enumitem} 
\usepackage{bm} 
\usepackage[bookmarks=false, draft]{hyperref}
\usepackage{times}
\usepackage{amssymb}
\usepackage{caption}
\usepackage{color}
\usepackage{array}
\usepackage{multirow}
\usepackage{mathtools}
\title{A Comparison of Audio Signal Preprocessing Methods \\for Deep Neural Networks on Music Tagging}
\twoauthors
  {Keunwoo Choi, Gy\"orgy Fazekas, Mark Sandler\sthanks{FAST IMPACt EPSRC Grant EP/L019981/1 and the European Commission H2020 research and innovation grant AudioCommons (688382). Mark Sandler acknowledges the support of the Royal Society as a recipient of a Wolfson Research Merit Award.} }
	{Queen Mary University of London, London, UK\\
	Centre for Digital Music, EECS\\
	E1 4FZ, London, UK\\
	\texttt{keunwoo.choi{@}qmul.ac.uk}
	}
  {Kyunghyun Cho\sthanks{Kyunghyun Cho thanks the support by Facebook, Google (Google Faculty Award 2016) and NVidia (GPU Center of Excellence 2015-2016)}}
	{New York University, NY, USA\\
	Center for Data Science\\
	715 Broadway, New York, NY, USA\\
	\texttt{kyunghyun.cho@nyu.edu}}

\begin{document}
\maketitle
\begin{abstract}
In this paper, we empirically investigate the effect of audio preprocessing on music tagging with deep neural networks. We perform comprehensive experiments involving audio preprocessing using different time-frequency representations, logarithmic magnitude compression, frequency weighting, and scaling. We show that many commonly used input preprocessing techniques are redundant except magnitude compression. 

\end{abstract}
\section{Introduction}

Many music information retrieval researches using deep learning usually focus on optimising the hyperparameters which specify the network structure. Meanwhile, the audio preprocessing stage is often heuristically decided and not subject to optimisation.

Although neural networks are known to be universal function approximators \cite{hornik1991approximation}, training efficiency and performance may vary significantly with different training methods as well as generic techniques including preprocessing the input data \cite{lecun2012efficient}. In other words, a neural network can \textit{represent} any function but it does not mean it can effectively \textit{learn} any function. Therefore, both empirical decisions and domain knowledge are crucial since choosing between various preprocessing methods can be seen as a non-differentiable choice function, therefore it cannot be optimised using gradient-based learning methods.

For example, mel-spectrograms have been preferred over short-time Fourier transform in many tasks \cite{choi2017tutorial} because it was considered to have enough information despite of its smaller size, i.e., efficient yet effective. When a time-frequency representation $\textbf{X}$ is given, one of the most common preprocessing approaches is to apply logarithmic compression, i.e., $\log(\textbf{X}+\alpha)$ where $\alpha$ can be arbitrary constants such as very small number (e.g. $10^{-7}$) or $1$. However, the performances of these methods are usually not strictly compared. 

In this paper, we focus on audio preprocessing strategies for deep convolutional neural networks for music tagging. By providing empirical results with various preprocessing strategies, we aim to demystify the effect of audio preprocessing methods on the performances. This will help researchers design deep learning systems for music research.

\vspace{-0.3cm}
\section{Experiments and Discussions}\label{sec:input_postp}
\begin{figure}[t]
	\centering
	\includegraphics[width=0.75\columnwidth]{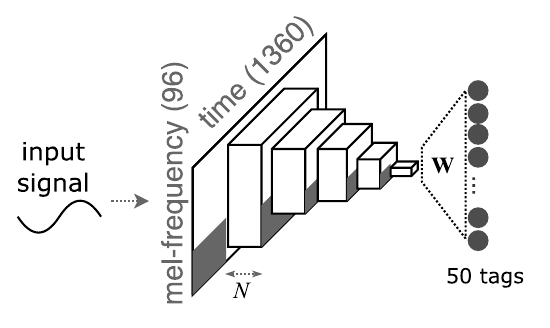}
	\caption{Network structure of the 5-layer ConvNet. $N$ refers to the number of feature maps (which is set to 32 for all layers in this paper) while $\textbf{W}$ refers to the weights matrix of the fully-connected output layer.)}
	\label{fig:transfer_diagram}
\end{figure}

\vspace{-0.15cm}
A representative network structure needs to be defined to compare the effects of audio preprocessing. A ConvNet (convolutional neural networks) with 2D kernels and 2D convolution axes was chosen. This showed a good performance with efficient training in a prior benchmark \cite{choi2017convolutional}, where the model we selected was denoted \textit{k2c2}, indicating 2D kernels and convolution axes. As illustrated in Figure \ref{fig:transfer_diagram}, homogeneous 2D ($3$$\times$$3$) convolutional kernels are used in every convolutional layer. The input has a single channel, 96 mel bins, and 1,360 temporal frames, denoted as (1,~96,~1360). The figures in Table \ref{table:c_ConvNet} denote the number of channels (N), kernel height and kernel width for convolutional layers and subsampling height, subsampling width for max-pooling layers. Here, the height and width corresponds to the frequency- and time-axes respectively. Exponential linear unit (ELU) is used as an activation function in all convolutional layers \cite{clevert2015fast}.

For the training of music tagger, we used the Million Song Dataset (MSD) \cite{bertin2011million} with preview audio clips. The training data are 30-second stereo mp3 files with a sampling rate of 22,050Hz and 64~kbps constant bit-rate encoding. For efficient training in our experiments, we downmix and downsample the signals to 12~kHz after decoding and trim the audio duration to 29-second to ensure equal-sized input signals. The short-time Fourier transform and melspectrogram are computed using a hop size of 256~samples (21.3~ms) with a 512-point discrete Fourier transform aggregated to yield 96 mel bins per frame. The preprocessing is performed using \textit{Librosa} \cite{mcfee_brian_2017_293021} and \textit{Kapre} \cite{choi2017kapre}. Total 224,242 tracks are used and split into train/validation/test sets, 201,672/12,633/28,537 tracks respectively.\footnote{The network implementation and split setting are online: \\\url{https://github.com/keunwoochoi/transfer_learning_music} and \\\url{https://github.com/keunwoochoi/MSD_split_for_tagging}}
During training, the binary cross-entropy function is used as a loss function. For the acceleration of stochastic gradient descent, we use adaptive optimisation based on Adam \cite{DBLP:journals/corr/KingmaB14}. The experiment is implemented in Python with \textit{Keras} \cite{chollet2015} and \textit{Theano} \cite{bastien2012theano} as deep learning frameworks.

In the all experiments, area under curve - of receiver operating characteristic (AUC) is used as a metric. Although it can be lower than 0.5 in theory, AUC practically ranges in [0.5, 1.0] as random and perfect predictions show 0.5 and 1.0 of AUC respectively. The reported AUC scores are all measured on the test set.

  	\vspace{-0.2cm}

\begin{table}[t]
	\centering
	\caption{Details of the ConvNet architecture shown in Figure \ref{fig:transfer_diagram}. 2-dimensional convolutional layer is specified by (channel, (kernel lengths in frequency, time)). Pooling layer is specified by (pooling length in frequency and time) }
	\begin{tabular}{|l|c|}
		\hline
		\multicolumn{2}{|c|}{\textit{\textbf{input (1, 96, 1360)}}} \\ \hline
		Conv2d and Max-Pooling         & (32, (3, 3)) and (2, 4)                 \\ \hline
		Conv2d and Max-Pooling         & (32, (3, 3)) and (4, 4)                 \\ \hline
		Conv2d and Max-Pooling         & (32, (3, 3)) and (4, 5)                 \\ \hline
		Conv2d and Max-Pooling         & (32, (3, 3)) and (2, 4)                 \\ \hline
		Conv2d and Max-Pooling         & (32, (3, 3)) and (4, 4)                 \\ \hline
        Fully-connected layer         & (50)                                    \\ \hline
		\multicolumn{2}{|c|}{\textit{\textbf{output (50)}}}  \\ \hline
	\end{tabular}
	\label{table:c_ConvNet}
\end{table} 

\subsection{Variance by different initialisation}

\begin{figure}[t]
	\centering
	\includegraphics[width=\columnwidth]{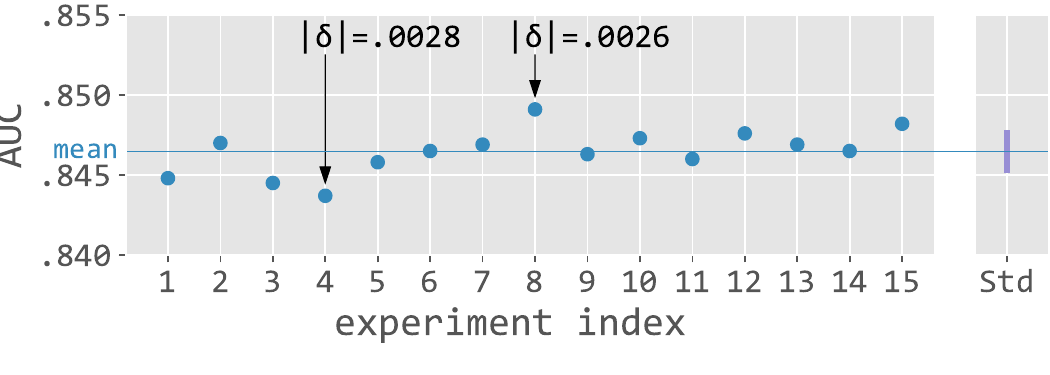}
	\caption{AUC scores (left), their mean, and the standard deviation of the AUC scores (right). The two deltas on the plot indicate the difference between the average AUC and the scores of experiments 4 and 8.}
	\label{fig:repeat_exp}
\end{figure}

In deep learning, using \textit{K}-fold cross-validation is not a standard practice for two reasons. First, with large enough data and a good split of train, validation and test sets, the model can be trained with small variance. Second, the cost of hyperparameter search is very high and it makes repeating experiments too expensive in practice. For these reasons, we do not cross-validate the ConvNet in this study. Instead, we present the results of repeated experiments with fixed network and training hyperparameters, such as training example sequences and batch size. This experiment therefore measures the variance of the model introduced by different weight initialisations of the convolutional layers. For this, a normal distribution is used following He~et~al.~\cite{he2015delving}, which has been shown to yield a stable training procedure. 

The results are summarised in Figure \ref{fig:repeat_exp}. This shows the AUC scores of 15 repeated experiments on the left as well as their standard deviation on the right. Small standard deviation indicates that we can obtain a reliable, precise score by repeating the same experiments for a sufficient number of times. 
The two largest differences observed between the average AUC score and that of experiment 4 and 8 (AUC differences of 0.0028 and 0.0026 respectively) indicate that we may obtain up to $\sim 0.005$ AUC difference among experiment instances. Based on this, we can assume that an AUC difference of $< 0.005$ is non-significant in this paper.

  	\vspace{-0.2cm}

\subsection{Time-frequency representations}\label{ssec:stft_mel}

STFT and melspectrogram have been the most popular input representations for music classification \cite{choi2017tutorial}. Although sample-based deep learning methods have been introduced, 2-dimensional representations would be still useful in the near future for efficient training. Melspectrograms provide an efficient and perceptually relevant representation compared to STFT \cite{moore2012introduction} and have been shown to perform well in various tasks \cite{van2013deep, dieleman2014end, schluter2014improved, ullrich2014boundary, choi2016automatic}. However, an STFT is closer to the original signal and neural networks may be able to learn a representation that is more optimal to the given task than mel-spectrograms. This requires large amounts of training data however, as reported in \cite{choi2016automatic} where using melspectrograms outperformed STFT with a smaller dataset. 

\begin{figure}[t]
	\centering
	\includegraphics[width=\columnwidth]{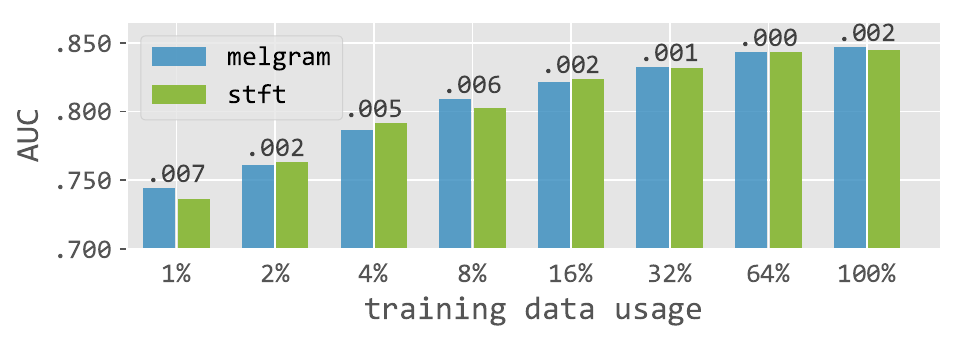}
	\caption{Performances of predictions with melspectrogram and STFT with varying training data sizes. The numbers above bars indicate the absolute performance differences between melspectrograms and STFTs.} 
	\label{fig:stft_mel_by_data}
\end{figure}

Figure \ref{fig:stft_mel_by_data} shows the AUC scores obtained using  $\log$(STFT) vs. $\log$(melspectrogram) while varying the size of the utilised training data. Although there are small differences on AUC scores up to 0.007, neither of them outperforms the other, especially when enough data is provided. This rebuts a previous result in \cite{choi2016automatic} because melspectrograms did not have a clear advantage here even with a small training data size. This may be due to the difference in frequency resolution of the representations used and summarised as follows.

\begin{itemize}
  \item STFT in \cite{choi2016automatic}: $6000/129$$=$$46.5$ Hz (256-point FFT with 12 kHz sampling rate)
  	\vspace{-0.2cm}
  \item STFT in our work: $6000/257$$=$$23.3$ Hz (512-point FFT with 12 kHz sampling rate)
    	\vspace{-0.2cm}
  \item Melspectrogram in \cite{choi2016automatic} and our work: $35.9$ Hz for frequency $<1$ kHz (96 mel-bins and by \cite{slaney1998auditory} and \cite{mcfee_brian_2017_293021})
\end{itemize}

In \cite{choi2016automatic}, the frequency resolution of the STFT was lower than that of the melspectrogram to enable comparing them with similar number of frequency bins. On the contrary, STFT of higher frequency resolution is used in our experiment. Nevertheless, it is found to be only as good as melspectrogram in terms of performance, i.e., the network does not take advantage of finer input. This means overall that, in practice, melspectrogram may be preferred since its smaller size leads to reduced computation in training and prediction. The figure also illustrates how much the data size affects the performance. Exponentially increasing data size merely results in a linear AUC improvement. AUC starts to converge at \texttt{64\%} and \texttt{100\%}.

\subsection{Analysis of scaling effects and frequency-axis weights} \label{ssec:freq_scale}

\begin{figure}[t]
	\centering
	\includegraphics[width=\columnwidth]{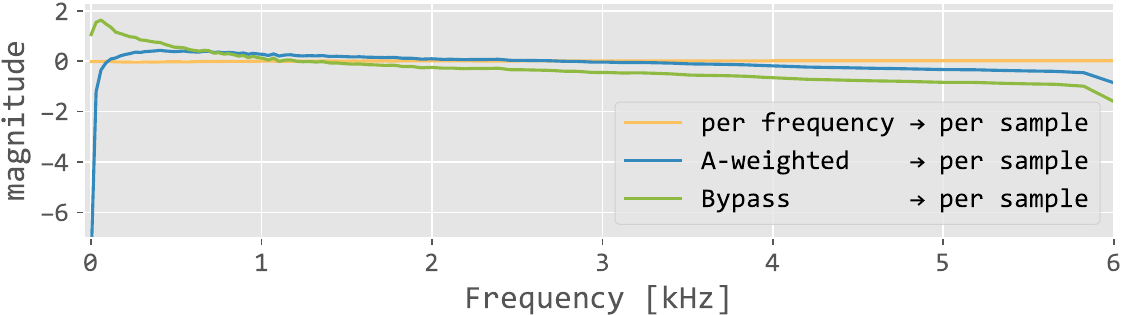}
	\caption{Average frequency magnitude of randomly selected 200 excerpts with three frequency-axis normalisation. A per-sample (excerpt) standardisation follows to remove the effect of different overall average magnitude.}
	\label{fig:preproc_freq_response}
\end{figure}

In this section, we discuss the effects of magnitude manipulation. Preliminary experiments suggested that there might be two independent aspects to investigate; {\em i)} frequency-axis weights and {\em ii)} magnitude scaling of each item in the training set. Examples of frequency-axis weights are illustrated in Figure \ref{fig:preproc_freq_response}, where different weighting schemes are plotted. Our experiment is designed to isolate these two effects. We tested two input representations \textit{log-melspectrogram} vs. \textit{melspectrogram}, with three frequency weighting schemes \textit{per-frequency}, \textit{A-weighting} and \textit{bypass}, as well as two scaling methods $\times 10$ (\texttt{on}) and $\times 1$ (\texttt{off}), yielding $2$$\times$$3$$\times$$2$$=$$12$ configurations in total. 
We summarise the mechanism of each block as follows. First, there are three frequency weights schemes. 
\begin{itemize}[leftmargin=*]
	\item \texttt{per-frequency stdd}: Often called spectral whitening. Compute means and standard deviations across time, i.e., per-frequency, and standardise each frequency band using these values. The average frequency response becomes flat (equalised). This method has been used in the literature for tagging \cite{choi2016automatic}, singing voice detection \cite{schluter2016learning} and transcription \cite{sigtia2015end}.
	\vspace{-0.2cm}
	\item \texttt{A-weighted}: Apply the international standard IEC 61672:2003 A-weighting curve, which approximates human perception of loudness as a function of frequency.
	\vspace{-0.5cm}
	\item \texttt{Bypass}: Do not apply any processing, i.e., $f:
	\mathbf{X} \to \mathbf{X}$
\end{itemize}

There are two other blocks which are mutually independent and also independent to frequency weights schemes.

\begin{itemize}[leftmargin=*]
	\item \texttt{per-sample stdd}: Excerpt-wise normalisation with its overall mean and standard deviation, i.e., using statistics across time and frequency of each spectrogram.
		\vspace{-0.15cm}
	\item \texttt{$\times$10 scaler}: Multiply the input spectrogram by 10, i.e.,\newline $f:
	\mathbf{X} \to 10$$\mathbf{X}$. 
\end{itemize}

\subsubsection{Frequency weighting} This process is related to the loudness, i.e., human perception of sound energy \cite{moore2012introduction}, which is a function of frequency. The sensitivity of the human auditory system drops substantially below a few hundred Hz\footnote{See equal-loudness contours e.g. in ISO 226:2003.}, hence music signals typically exhibit higher energy in the lower range to compensate for the attenuation. This is illustrated in Figure \ref{fig:preproc_freq_response}, where uncompensated average energy measurements corresponding to the \textit{Bypass} curve (shown in green) yield a peak at low frequencies. This imbalance affects neural network activations in the first layer which may influence performance. To assess this effect, we tested three frequency weighting approaches. Their typical profiles are shown in Figure \ref{fig:preproc_freq_response}. In all three strategies, excerpt-wise standardisation is used to alleviate scaling effects (see Section \ref{sssec:scale_fx}).

\begin{figure}[t]
	\centering
	\includegraphics[width=\columnwidth]{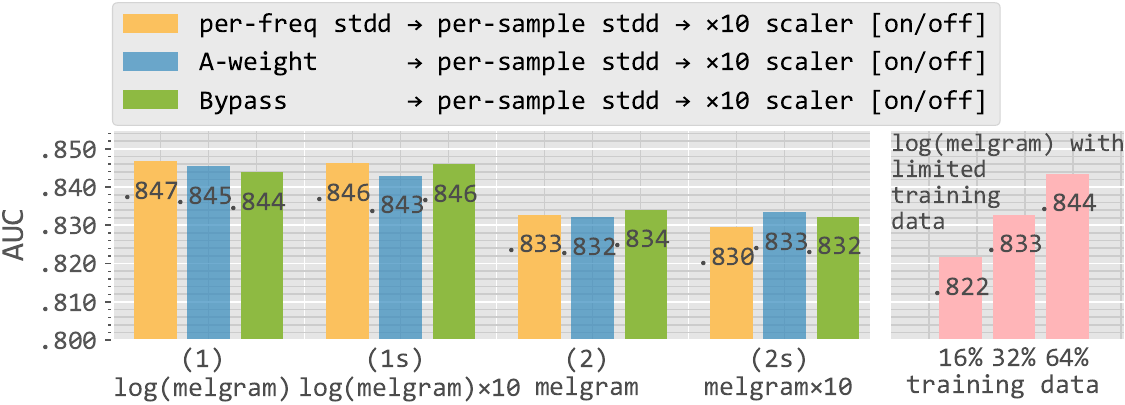}
	\caption{Performance comparisons of different preprocessing procedures. From left to right, four groups of scores are from different magnitude processing (melspectrogram in decibel scale and linear scale), with additional $\times10$scaler turned on/off. In each group, yellow/blue/green bars indicates different frequency-axis processing as annotated in the legend. Logarithmic compression is done before other preprocessings are applied. }
	\label{fig:result_preproc_all}
	
	\vspace{0.3cm}
	  	
	\centering
	\includegraphics[width=\columnwidth]{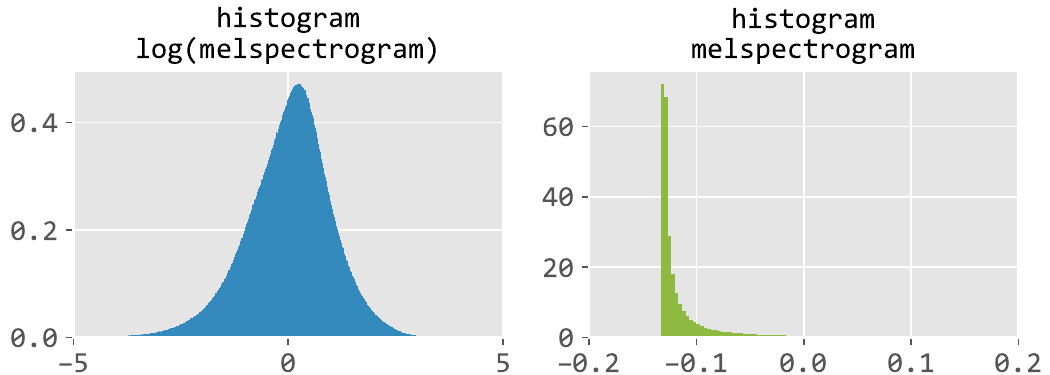}
	\caption{histograms of the magnitude of melspectrogram time-frequency bins with (left) and without (right) logarithmic compression. The number of bins are 100 and both are normalised, i.e., $
		\sum_{i=1}^{100} 0.01 \times y_i=1$. Log compression significantly affects the histogram, making the distribution Gaussian (left), otherwise extremely skewed (right). This is after standardisation and based on randomly selected 100 tracks from the training set.}
	\label{fig:log_mag_hist}
\end{figure}

Our test results show that networks using the three strategies all achieve similar AUC scores. The performance differences within four groups, \{\texttt{1}, \texttt{1s}, \texttt{2}, \texttt{2s}\} in Figure \ref{fig:result_preproc_all} are small and none of them are governing the others. The curves in Figure \ref{fig:preproc_freq_response} show the average input magnitudes over frequency. These offsets change along frequency, but the change does not seem large enough to corrupt the local patterns due to the locality of ConvNets, and therefore the network is learning useful representations without significant performance differences within each group.

\vspace{-0.2cm}

\subsubsection{Analysis of scaling effects} \label{sssec:scale_fx}
\vspace{-0.1cm}

We may assume a performance increase if we scale the overall magnitudes for a number of reasons. During training using gradient descent, the gradient of error with respect to weights $\frac{\partial E}{\partial W}$ is proportional to $\frac{\partial}{\partial W} f(W^\top X)$ where $f$ is the activation function. This means that the learning rate of a layer is proportional to the magnitude of input $X$. In particular, the first layer usually receives the weakest error backpropagation, hence scaling of the input may affect the overall performance.

We tested the effect of this with the results shown in Fig. \ref{fig:result_preproc_all}. To this end, consider comparing the same-coloured bars of \{\texttt{1} vs. \texttt{1s}\} and \{\texttt{2} vs. \texttt{2s}\}. Here, the scaling factor is set to 10 for illustration, however many possible values $<$100 were tested and showed similar results. In summary, this hypothesis is rebutted -- scaling did not affect the performance. 
The analysis of trained weights revealed that different magnitudes of the input only affects the bias of the first convolutional layer. Training with scaling set to $\times 10$ results in 3.4 times larger mean absolute value of the biases in the first layer. This is due to batch normalization \cite{ioffe2015batch} which compensates for the different magnitudes by normalizing the activations of convolutional layers.

\subsection{Log-compression of magnitudes}\label{ssec:log}

Lastly, we discuss how logarithmic compression of magnitudes, i.e. decibel scaling, affects performance. This is considered standard preprocessing in music information retrieval. The procedure is motivated by the human perception of loudness \cite{moore2012introduction} which has logarithmic relationship with the physical energy of sound. 
Although learning a logarithmic function is a trivial task for neural networks, it can be difficult to implicitly learn an optimal nonlinear compression when it is embedded in a complicated task. A nonlinear compression was also shown to affect the performance in visual image recognition using neural networks \cite{DBLP:journals/corr/DodgeK16}. Figure~\ref{fig:log_mag_hist} compares the histograms of the magnitudes of time-frequency bins after zero-mean unit-variance standardisation. On the left, a logarithmically compressed melspectrogram shows an approximately Gaussian distribution without any extreme values. Meanwhile, the bins of linear melspectrogram on the right is extremely condensed in a very small range while they range in wider region overall. This means the network should be trained with higher numerical precision to the input, hence more vulnerable to noise.

As a result, decibel-scaled melspectrograms always outperform the linear versions as shown in Fig \ref{fig:result_preproc_all}, where the same-coloured bars should be compared across within \{\texttt{1}~vs.~\texttt{2}\} and \{\texttt{1s}~vs.~\texttt{2s}\}. Colours indicate normalization schemes while \{\texttt{1}~vs.~\texttt{1s}\} and \{\texttt{2}~vs.~\texttt{2s}\} compare  effect, both of which are explained in Section~\ref{ssec:freq_scale}\footnote{Decibel-scaled STFT also outperformed linear STFT in our unreported experiments.}. Compared to the performance differences while controlling the training set size (the pink bar charts on the right of Figure \ref{fig:result_preproc_all}) the additional work introduced by not using decibel scaling can be roughly estimated by comparing these scores to those networks when the training data size is limited (in pink). While this also depends on other configurations of the task, seemingly twice the data is required to compensate for the disadvantage of not using a decibel scaled representation.

\section{Conclusion}
In this paper, we have shown that some of the input preprocessing methods can affect the performance. We quantify this in terms of the size of the training data required to achieve similar performances. Among several preprocessing techniques tested in this study, only logarithmic scaling of the magnitude resulted in significant improvement. In other words, the network was resilent to most modifications of the input data except logarithmic compression of magnitudes in various time-frequency representations. Although we focused on the music tagging task, our results provide general knowledge applicable in many similar machine-listening problems, e.g., music genre classification or the prediction of environmental sound descriptors. 

\bibliographystyle{IEEEtran}
\bibliography{convnets}

\end{document}